\documentclass[epj]{svjour}
\usepackage{graphicx}
\newcommand{\SM}{S\&M}
\newcommand{\al}{$\alpha$}
\newcommand{\g}{$\gamma$}
\newcommand{\gamal}{$\gamma_\alpha^2$}
\newcommand{\thal}{$\theta_\alpha^2$}

\newcommand{\raa}{($\alpha$,$\alpha$)}

\newcommand{\rag}{($\alpha$,$\gamma$)}

\newcommand{\tiii}{$^{42}$Ti}
\newcommand{\tiiv}{$^{44}$Ti}
\newcommand{\tinull}{$^{50}$Ti}
\newcommand{\canull}{$^{40}$Ca}
\newcommand{\crvi}{$^{46}$Cr}
\newcommand{\criv}{$^{54}$Cr}
\begin{document}
\title{
$\alpha$-cluster states in $^{46,54}$Cr from double-folding potentials
}
%
%
\author{Peter Mohr\inst{1,2}
\thanks{\emph{Email: WidmaierMohr@t-online.de; mohr@atomki.mta.hu}} 
}                     
%
%
\institute{
Diakonie-Klinikum, D-74523 Schw\"abisch Hall, Germany
 \and 
Institute for Nuclear Research (Atomki), H-4001 Debrecen, Hungary}
\date{Received: date / Revised version: date}
%
\abstract{
$\alpha$--cluster states in $^{46}$Cr and $^{54}$Cr are investigated in the
double-folding model. This study complements a recent similar work of Souza
and Miyake \cite{Sou17} which was based on a specially shaped
potential. Excitation energies, reduced widths, intercluster separations, and
intra-band transition strengths are calculated and compared to experimental
values for the ground state bands in $^{46}$Cr and $^{54}$Cr. The
$\alpha$-cluster potential is also applied to elastic scattering at low and
intermediate energies. Here, as a byproduct, a larger radial extent of the
neutron density in $^{50}$Ti is found.
\PACS{
      {21.60.Gx}{Cluster models}   \and
      {27.40.+z}{39 $\le$ A $\le$ 58}
     } 
} 
\maketitle
\section{Introduction}
\label{sec:intro}
\al -clustering is a very well-known phenomenon in nuclear physics which is
found in many nuclei across the chart of nuclides \cite{Hor12}. A lot of work
has been done for doubly-magic cores, i.e.\ $^{212}$Po = $^{208}$Pb $\otimes$
\al , $^{44}$Ti = $^{40}$Ca $\otimes$ \al , $^{20}$Ne = $^{16}$O $\otimes$ \al
, and $^{8}$Be = $^{4}$He $\otimes$ \al . The wide gap between $^{212}$Po and
$^{44}$Ti was often filled by studies of nuclei with semi-magic ($N =50$)
cores like $^{96}$Ru = $^{92}$Mo $\otimes$ \al\ or $^{94}$Mo = $^{90}$Zr
$\otimes$ \al . A detailed introduction into the nuclear cluster model is
provided in a dedicated special issue of Prog.\ Theor.\ Phys.\ Suppl.\ {\bf
  132}, \cite{Ohk98,Mic98,Yam98,Sak98,Ueg98,Has98,Koh98,Toh98}.  Very
recently, Souza and Miyake \cite{Sou17} (hereafter: \SM ) have extended these
studies towards the chromium isotopes. From the $Q_A/A_T$ systematics they
have identified $^{46,54}$Cr = $^{42,50}$Ti $\otimes$ \al\ as preferred nuclei
for \al -clustering. Their subsequent study finds that \crvi\ has a
significant degree of \al -clustering, whereas the reduced \al\ widths in
\criv\ are about a factor of three lower. Interestingly, earlier studies above
$^{44}$Ti have focused on $^{48}$Cr which is considered as a $^{40}$Ca core
plus two \al\ particles \cite{Des02,Sak02}.

The present study is motivated as an extension of the work of \SM . In the
approach by \SM\ it is first intended to find a two-body potential which is
able to reproduce the energies of \al -cluster states in \crvi\ and \criv
. This potential is then used to calculate the (quasi-)bound state wave
functions $u(r)$ and to derive reduced widths \gamal\ and \thal\ and
transition strengths $B(E2)$.

The approach of the present study, i.e., the \al -cluster model in combination
with double-folding potentials, has been widely used for the above mentioned
doubly-magic cores. Properties of $^{8}$Be = $^{4}$He $\otimes$ \al\ and the
$^{4}$He\rag $^{8}$Be capture cross section were calculated in
\cite{Mohr93}. A detailed study of $^{19}$F = $^{15}$N $\otimes$ \al\ and
$^{20}$Ne = $^{16}$O $\otimes$ \al\ is given in \cite{Abe93} which was later
extended to \rag\ capture reactions \cite{Wil02,Mohr05,Mohr06}. $^{44}$Ti =
$^{40}$Ca $\otimes$ \al\ and $^{40}$Ca\raa $^{40}$Ca elastic scattering was
studied in \cite{Atz96}, and the mass region around $A \approx 40$ was also
reviewed in detail in \cite{Sak98}.  $^{212}$Po = $^{208}$Pb $\otimes$ \al ,
the \al - and \g -decay properties of $^{212}$Po, and \al -elastic scattering
are investigated in \cite{Hoy94,Ohk95}, and a study for $^{104}$Te =
$^{100}$Sn $\otimes$ \al\ is given in \cite{Mohr07}. A series of $N = 50$
$\otimes$ \al\ nuclei were investigated in \cite{Ohk95,Ohk09,Mohr08}. Finally,
\al -decay properties of nuclei, including superheavy nuclei up to $A \approx
300$, are often described within the folding potential approach (e.g.,
\cite{Mohr00,Xu06,Mohr17}).

Here I focus on \crvi\ and \criv\ and compare the results of \SM\ to the
results from systematic double-folding potentials. Contrary to the approach of
\SM\ who are able to describe the excitation energies of all states in the
ground state band using a specially shaped potential, the double-folding
potentials of the present study require a minor $L$-dependent adjustment to
each state under study. This disadvantage of the folding potential is
compensated by two advantages: ($i$) The folding potentials can describe not
only bound state properties, but also elastic scattering over a wide energy
range. ($ii$) The typically smooth variation of the parameters of the folding
potentials can be used to identify peculiar properties of the nuclei under
study. Consequently, the approaches of \SM\ and this work should be considered
as complementary.

\section{Folding potential model}
\label{sec:fold}
In the present study the interaction between the \al\ particle and the core is
calculated from the folding procedure with the widely used energy- and
density-dependent DDM3Y interaction $v_{\rm{eff}}$:
\begin{equation}
V_F(r) = 
	\int \int \, \rho_P(r_P) \,\rho_T(r_T) \,
        v_{\rm{eff}}(s,\rho,E_{\rm{NN}}) \; d^3r_P \; d^3r_T
\label{eq:fold}
\end{equation}
For details of the folding approach, see e.g.\ \cite{Sat79,Kob84,Mohr13}.

The densities $\rho_P$ and $\rho_T$ of projectile and target are usually
derived from the experimental charge density distributions which are measured
by electron scattering and summarized in \cite{Vri87}. However, in the
present study, the density of \tiii\ for the \tiii -\al\ potential is not
available because \tiii\ is unstable and experimentally not yet accessible for
electron scattering (although electron scattering for unstable nuclei may
become feasible in the near future at the SCRIT facility \cite{Tsu17}). In
addition, the charge density distribution of \tinull\ for the \tinull
-\al\ potential may deviate from the matter density because the $N/Z$ ratio
deviates significantly from unity. Therefore, the present work uses
theoretical densities as provided as a part of the widely used statistical
model code TALYS \cite{TALYS}.

The following tests have been made to verify the theoretical density
distributions of TALYS. First, the calculations of \al -cluster states in
\tiiv\ of our previous work \cite{Atz96} were repeated using the theoretical
density of \canull\ from TALYS instead of an experimental density, and only
tiny deviations to the results in Tables II and III of \cite{Atz96} were
found. Second, folding potentials were calculated for \tinull -\al\ using
either the theoretical density from TALYS or a matter density derived from the
experimental charge density \cite{Vri87}. Both potentials were applied to
\tinull \raa \tinull\ elastic scattering at 25 MeV \cite{Gub81}, and it was
found that a better description of the experimental data was obtained using
the theoretical \tinull\ density distribution. Further details on elastic
scattering will be discussed later in Sect.~\ref{sec:elast}, including
scattering data at low and intermediate energies
\cite{Rob78,Pesl83}. Consequently, after these successful tests, the
theoretical density distributions from TALYS were used in the present study.

The total interaction potential $V(r)$ is given by 
\begin{equation}
V(r) = V_N(r) + V_C(r) = \lambda \, V_F(r) + V_C(r)
\label{eq:vtot}
\end{equation}
where the nuclear potential $V_N$ is the double-folding potential $V_F$ of
Eq.~(\ref{eq:fold}) multiplied by a strength parameter $\lambda \approx 1.1 -
1.3$ \cite{Atz96,Mohr13}. $V_C$ is the Coulomb potential in the usual form of
a homogeneously charged sphere with the Coulomb radius $R_C$ chosen the same
as the $rms$ radius of the folding potential $V_F$.

The strength parameter $\lambda$ is adjusted to reproduce the energies of the
bound states with $E < 0$ and quasi-bound states with $E > 0$ where $E = 0$
corresponds to the threshold of \al\ emission in the compound
nucleus. The number of nodes $N$ of the bound state
wave function was taken from the Wildermuth condition
\begin{equation}
Q = 2N + L = \sum_{i=1}^4 (2n_i + l_i) = \sum_{i=1}^4 q_i
\label{eq:wild}
\end{equation}
where $Q$ is the number of oscillator quanta, $N$ is the number of nodes, and
$L$ is the relative angular momentum of the $\alpha$-core wave function. $q_i
= 2n_i + l_i$ are the corresponding quantum numbers of the nucleons in the
$\alpha$ cluster. For the ground state bands of the nuclei in the lower
$fp$-shell I use $q_i = 3$, resulting in $Q = 12$, which leads to seven states
with $J^\pi$ from $0^+$ to $12^+$ for the even-even nuclei under study. This
choice is similar to \SM .

Typically, a smooth decrease of the potential strength parameter
$\lambda$ is found with increasing excitation energy or increasing angular
momentum \cite{Abe93,Mohr08}. For intermediate mass nuclei around $N = 50$ an
almost linear decrease of $\lambda$ is found for the whole ground state band
\cite{Mohr08} whereas for lighter nuclei the decreasing trend of $\lambda$
changes to an increasing $\lambda$ for states above $L \approx 6$
\cite{Abe93}.

The formalism for the calculations has been provided in earlier work
\cite{Abe93,Atz96,Mohr07,Mohr08}. Reduced widths $\gamma_\alpha^2$ are
determined using the same method as in \SM .

\section{Results and Discussion}
\label{sec:res}
\begin{figure}[htb]
  \includegraphics[width=\columnwidth]{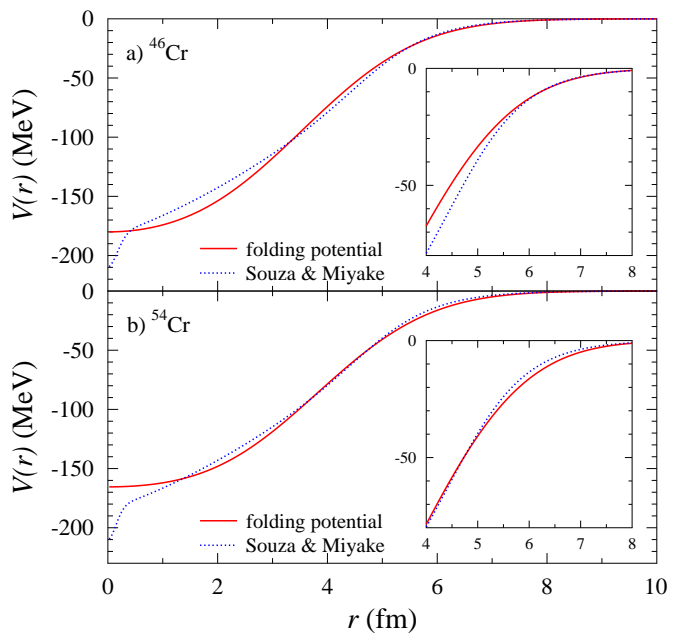}
\caption{
Comparison between the folding potential $\lambda \times V_F(r)$ and the
specially shaped potential of \SM\ for \crvi\ (upper) and \criv\ (lower). The
values for $\lambda$ are taken from the ground states (see Tables
\ref{tab:res_vi} and \ref{tab:res_iv}). The insets enlarge the region of the
nuclear surface which is most relevant for elastic scattering.
}
\label{fig:pot}
\end{figure}

The results for \crvi\ and \criv\ are summarized in Tables \ref{tab:res_vi}
and \ref{tab:res_iv}. For comparison, also the results of \SM\ are listed in
Tables \ref{tab:res_vi} and \ref{tab:res_iv}. The potentials are shown in
Fig.~\ref{fig:pot}.
\begin{table*}
\caption{\al -cluster properties of \crvi . For comparison, the second line
  for each state shows the results of \SM\ \cite{Sou17} . Experimental data
  have been taken from the ENSDF database \cite{ENSDF} which is based on
  \cite{NDS46} for \crvi . 
}
\label{tab:res_vi}
\begin{center}
\begin{tabular}{crrccrrrrrrr}
\hline
$J^\pi$ 
& \multicolumn{1}{c}{$E^\ast$} 
& \multicolumn{1}{c}{$E$} 
& $N$ & $L$ 
& \multicolumn{1}{c}{$\lambda$}
& \multicolumn{1}{c}{$J_R$} 
& \multicolumn{1}{c}{$\langle R^{2} \rangle^{1/2}$}
& \multicolumn{1}{c}{$\gamma_\alpha^2$}
& \multicolumn{1}{c}{$\theta_\alpha^2$}
& \multicolumn{1}{c}{$B(E2,L\rightarrow L-2)$}
& \multicolumn{1}{c}{$B(E2)_{\rm{exp}}$} \\
& (keV)
& (keV)
& & 
& 
& \multicolumn{1}{c}{(MeV\,fm$^3$)} 
& \multicolumn{1}{c}{(fm)}
& \multicolumn{1}{c}{(keV)}
& \multicolumn{1}{c}{(\%)}
& \multicolumn{1}{c}{(W.u.)}
& \multicolumn{1}{c}{(W.u.)} \\
%
\hline
%
$0^+$   &       $0.0$   &  $-6793.8$  &  6  &   0 & 1.2330  & 355.10 
&  4.416  & 2.607  & 0.843  & -- \\
& & & & & &
&  4.339  & 2.179  & 0.692  & -- \\
$2^+$   &     $892.2$   &  $-5901.6$  &  5  &   2 & 1.2241  & 352.54 
&  4.416  & 2.589  & 0.822  & 10.3 & 19(4)\\
& & & & & &
&  4.341  & 2.213  & 0.702  & 9.7 \\
$4^+$   &    $1987.1$   &  $-4806.7$  &  4  &   4 & 1.2204  & 351.48 
&  4.367  & 1.937  & 0.615  & 13.9 & \\
& & & & & &
&  4.299  & 1.690  & 0.536  & 13.0 \\
$6^+$   &    $3226.9$   &  $-3566.9$  &  3  &   6 & 1.2232  & 352.28 
&  4.270  & 1.037  & 0.329  & 13.2 & \\
& & & & & &
&  4.219  & 0.935  & 0.297  & 12.5 \\
$8^+$   &    $4817.4$   &  $-1976.4$  &  2  &   8 & 1.2293  & 354.04 
&  4.137  & 0.392  & 0.127  & 10.7 & \\
& & & & & &
&  4.121  & 0.374  & 0.119  & 10.3 \\
$10^+$  &    $6179.5$   &   $-614.3$  &  1  &  10 & 1.2479  & 359.40 
&  3.959  & 0.079  & 0.025  & 7.2 & \\
& & & & & &
&  4.010  & 0.084  & 0.027  & 7.1 \\
$12^+$  &    $8162.5$   &  $+1368.7$  &  0  &  12 & 1.2657  & 364.52 
&  3.766  & 0.008  & 0.003  & 3.6 & \\
& & & & & &
&  3.933  & 0.010  & 0.003  & 3.7 \\
%
\hline
\end{tabular}
\end{center}
\end{table*}
\begin{table*}
\caption{\al -cluster properties of \criv . For comparison, the second line
  for each state shows the results of \SM\ \cite{Sou17} . Experimental data
  have been taken from the ENSDF database \cite{ENSDF} which is based on
  \cite{NDS54} for \criv . 
}
\label{tab:res_iv}
\begin{center}
\begin{tabular}{crrccrrrrrrr}
\hline
$J^\pi$ 
& \multicolumn{1}{c}{$E^\ast$} 
& \multicolumn{1}{c}{$E$} 
& $N$ & $L$ 
& \multicolumn{1}{c}{$\lambda$}
& \multicolumn{1}{c}{$J_R$} 
& \multicolumn{1}{c}{$\langle R^{2} \rangle^{1/2}$}
& \multicolumn{1}{c}{$\gamma_\alpha^2$}
& \multicolumn{1}{c}{$\theta_\alpha^2$}
& \multicolumn{1}{c}{$B(E2,L\rightarrow L-2)$}
& \multicolumn{1}{c}{$B(E2)_{\rm{exp}}$} \\
& (keV)
& (keV)
& & 
& 
& \multicolumn{1}{c}{(MeV\,fm$^3$)} 
& \multicolumn{1}{c}{(fm)}
& \multicolumn{1}{c}{(keV)}
& \multicolumn{1}{c}{(\%)}
& \multicolumn{1}{c}{(W.u.)}
& \multicolumn{1}{c}{(W.u.)} \\
%
\hline
%
$0^+$   &       $0.0$   &  $-7927.9$  &  6  &   0 & 1.1213  & 315.54 
&  4.438  & 1.133  & 0.392  & -- \\
& & & & & &
&  4.290  & 0.631  & 0.218  & -- \\
$2^+$   &     $834.9$   &  $-7093.0$  &  5  &   2 & 1.1121  & 312.95 
&  4.438  & 1.117  & 0.386  & 8.5 & 14.4(6)\\
& & & & & &
&  4.290  & 0.637  & 0.221  & 7.5 \\
$4^+$   &    $1823.9$   &  $-6104.0$  &  4  &   4 & 1.1058  & 311.18 
&  4.395  & 0.821  & 0.284  & 11.5 & 26(9)\\
& & & & & &
&  4.249  & 0.471  & 0.163  & 10.0 \\
$6^+$   &    $3222.5$   &  $-4705.4$  &  3  &   6 & 1.0990  & 309.27 
&  4.320  & 0.465  & 0.161  & 11.0 & 18(5)\\
& & & & & &
&  4.181  & 0.268  & 0.093  & 9.7 \\
$8^+$   &    $4681.5$   &  $-3246.4$  &  2  &   8 & 1.0976  & 308.87 
&  4.202  & 0.173  & 0.060  & 9.0 & 12.8(17)\\
& & & & & &
&  4.089  & 0.101  & 0.035  & 8.0 \\
$10^+$  &    $6726.2$   &  $-1201.7$  &  1  &  10 & 1.0946  & 308.03 
&  4.062  & 0.043  & 0.015  & 6.3 & \\
& & & & & &
&  4.003  & 0.027  & 0.009  & 5.7 \\
$12^+$  &    $8825.4$   &   $+897.5$  &  0  &  12 & 1.0980  & 308.98 
&  3.888  & 0.005  & 0.002  & 3.2 & \\
& & & & & &
&  3.933  & 0.003  & 0.001  & 3.0 \\
%
\hline
\end{tabular}
\end{center}
\end{table*}
In general, good agreement is found between the present calculations in the
folding model and the previous results by \SM\ using a specially shaped
potential which was optimized for the reproduction of excitation
energies. This specially shaped potential of \SM\ is composed of a Woods-Saxon
(WS) plus a cubed WS$^3$ potential (as originally suggested in \cite{Buck95})
which is further modified by a (1+Gaussian) term at small
radii. Interestingly, it turns out that this specially shaped potential
becomes very similar to the folding potential of the present study except at
very small radii (see Fig.~\ref{fig:pot}). Thus, the good agreement between
the calculations is not surprising. For completeness it may be noted that
specially shaped Gaussian-modified Woods-Saxon potentials have also been
applied successfully to elastic scattering e.g.\ in \cite{Mic83,Mic00}.

The main difference between the folding potential of the present study and the
potential by \SM\ is the additional Gaussian dip at very small radii in the
\SM\ potential. This dip mainly reduces the energy of the $0^+$ ground state
whereas states with $L > 0$ are practically not affected because of the
centrifugal barrier which scales with $L(L+1)/r^2$ and thus dominates at small
radii $r$ (see also Fig.~3 of \SM ).

\begin{figure}[htb]
  \includegraphics[width=\columnwidth]{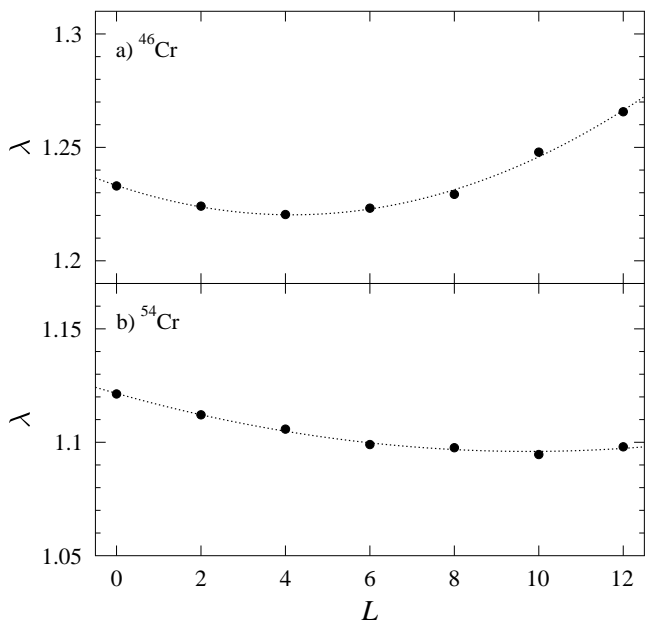}
\caption{
Variation of the potential strength parameter $\lambda$ as a function of
$L$. Minor variations for $\lambda$ are found for \crvi\ (below 4\%, upper
part) and and for \criv\ (below 2\%, lower part). The lines are quadratic
fits.
}
\label{fig:lambda}
\end{figure}

\SM\ attempt to reproduce the excitation energies of all states within the
yrast band by the choice of a specially shaped potential. Contrary to that
approach, the folding potential in the present study requires slight
adjustments of the strength parameter $\lambda$ to reproduce the energies of
the (quasi-)bound states under study. Strength parameters $\lambda$ of about
1.22 to 1.27 are found for \crvi , and $\lambda \approx 1.10 - 1.12$ is
obtained for \criv . The variations of $\lambda$ remain very small (below 4\%
for \crvi\ and below 2\% for \criv ). A smooth dependence of the strength
parameter $\lambda$ on the angular momentum $L$ of the bound state is found
(see Fig.~\ref{fig:lambda}). However, there is a significant difference
between \crvi\ and \criv . For \criv\ there is a smooth decrease of $\lambda$,
and all $\lambda$ values for $L \geq 6$ are practically constant within less
than 1\%. For \crvi\ a clear increase of $\lambda$ is found for $L > 6$. A
similarly increasing $\lambda$ for large $L$ was also found for \al -cluster
states in \tiiv\ \cite{Atz96}.

The $L$ dependence of $\lambda$ can be fitted by a parabola
\begin{equation}
\lambda(L) = \lambda_0 + \Delta \lambda \times (L - L_0)^2
\label{eq:lambda}
\end{equation}
with the values $\lambda_0 = 1.22027$ (1.09596), $\Delta \lambda = 7.4881
\times 10^{-4}$ ($2.6875 \times 10^{-4}$), and $L_0 = 4.1566$ (9.7575) for
\crvi\ (\criv ). The deviation between the parabolic fit in
Eq.~(\ref{eq:lambda}) and the $\lambda$ values in Tables \ref{tab:res_vi} and
\ref{tab:res_iv} is typically of the order of 0.001 or below, corresponding to
energy shifts of less than 100 keV. Thus, the experimental excitation energies
of the yrast bands in \crvi\ and \criv\ can be nicely reproduced in the
folding potential model in combination with a three-parameter parabolic fit
for the potential strength parameter $\lambda$.

\begin{figure}[htb]
  \includegraphics[width=\columnwidth]{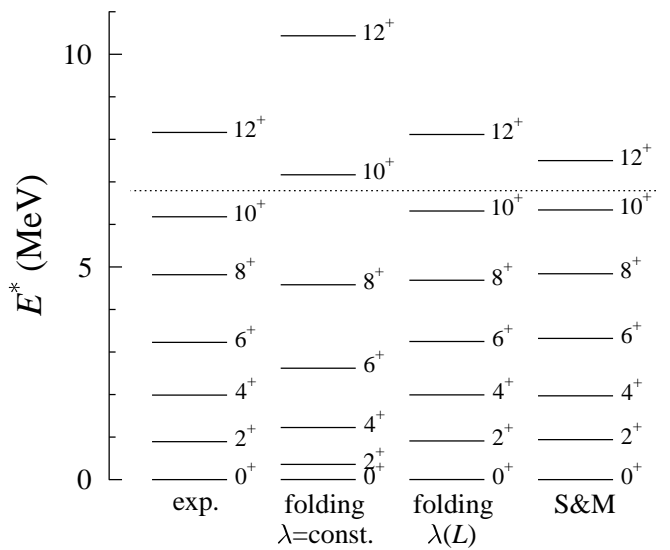}
\caption{
Experimental and calculated level scheme of \crvi . The dashed horizontal line
indicates the \al\ threshold. The level scheme has been calculated from the
folding potential with a fixed strength parameter $\lambda = 1.233$ and using
$\lambda(L)$ from Eq.~(\ref{eq:lambda}). The recent results of \SM\ are also
shown.
}
\label{fig:cr46level}
\end{figure}

\begin{figure}[htb]
  \includegraphics[width=\columnwidth]{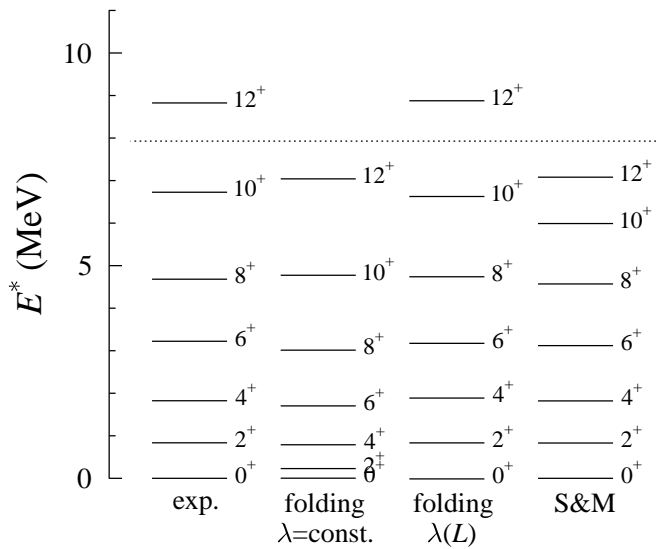}
\caption{
Experimental and calculated level scheme of \criv . The dashed horizontal line
indicates the \al\ threshold. The level scheme has been calculated from the
folding potential with a fixed strength parameter $\lambda = 1.121$ and using
$\lambda(L)$ from Eq.~(\ref{eq:lambda}). The recent results of \SM\ are also
shown.
}
\label{fig:cr54level}
\end{figure}

The experimental and calculated level schemes of \crvi\ and \criv\ are shown
in Figs.~\ref{fig:cr46level} and \ref{fig:cr54level}. The folding potential
with constant strength ($\lambda$ adjusted to the $0^+$ ground states)
shows a compressed rotational spectrum, in particular for states with small
$L$. A perfect reproduction of the excitation energies is obtained using the
slightly $L$-dependent strength parameter $\lambda(L)$ in
Eq.~(\ref{eq:lambda}). The potential by \SM\ describes the excitation
energies, too.

For completeness it has to be noted that the reproduction of excitation
energies (as shown in Figs.~\ref{fig:cr46level} and \ref{fig:cr54level}) is
not trivial. Usual Woods-Saxon potentials show the trend of spectral
inversion, i.e.\ excited states with $J > 0$ are located below the $0^+$
ground state (e.g., \cite{Mohr08}). Using typical parameters like $R = 1.3$ fm
and $a = 0.65$ fm for \crvi\ (\criv ), the first $2^+$ state is found at
$-0.16$ MeV ($-0.22$ MeV), and the $12^+$ state appears at $-8.62$ MeV
($-9.88$ MeV). The situation becomes even worse as soon as Woods-Saxon
parameters are chosen which were adjusted to elastic scattering. E.g., the
geometry of the widely used potential by McFadden and Satchler \cite{McF66}
($R = 1.4$ fm, $a = 0.52$ fm) leads to excitation energies of the $2^+$ state
of $-0.36$ MeV ($-0.40$ MeV) for \crvi\ (\criv ). Thus, results from usual
Woods-Saxon potentials are not included in Figs.~\ref{fig:cr46level} and
\ref{fig:cr54level}.

As soon as the potential strength is determined by adjustment of the strength
parameter $\lambda$, the corresponding wave functions $u(r)$ can be calculated
by numerical solution of the Schr\"odinger equation. Reduced widths
$\gamma_\alpha^2$ and $\theta_\alpha^2$, $rms$ intercluster separations
$\langle R^2\rangle^{1/2}$, and $B(E2)$ transition strengths result directly
from these wave functions (similar to \SM ). Qualitatively, the general
findings of \SM\ are reproduced in the present study: \\
\noindent ($i$) The $rms$ intercluster separation decreases slightly with
increasing angular momentum $L$. \\
\noindent ($ii$) The reduced widths $\gamma_\alpha^2$ and $\theta_\alpha^2$
decrease significantly with increasing $L$. \\
\noindent ($iii$) \al -clustering is more pronounced in \crvi\ compared to
\criv .\\
\noindent
Similar results have been obtained in earlier studies for the neighboring
nucleus $^{44}$Ti and also for heavier nuclei (e.g.,
\cite{Sou15,Ohk95,Mic88,Buck75}). 

However, there are also some differences between the work of \SM\ and the
present study. The $rms$ intercluster separations are slightly larger in the
present double-folding model. This holds in particular for states with small
angular momentum $L$ where smaller values in \SM\ result from the Gaussian dip
at small radii. In the present work the reduced widths $\theta_\alpha^2$ are
about 20\% higher for states with low $L$ in \crvi\ and almost a factor of two
higher for \criv . $B(E2)$ transition strengths are also slightly higher (of
the order of 10\%) in the present study. Furthermore, it should be pointed out
that the calculated $B(E2)$ transition strengths agree reasonably well with
the experimental data (whenever available). Similar to the work of \SM , no
effective charges are required here.

The $rms$ intercluster separations require a special discussion. Typically,
strong \al -cluster states are characterized by large $rms$ intercluster
separations $\langle R^2 \rangle^{1/2}$ (as also found by \SM ). The present
results in Tables \ref{tab:res_vi} and \ref{tab:res_iv} indicate that the
$\langle R^2 \rangle^{1/2}$ are close and even slightly larger for \criv\ with
its smaller reduced widths $\theta_\alpha^2$. However, as soon as the radii
$\langle R^2 \rangle^{1/2}$ are normalized to the radius of the compound
radius by $A_C^{1/3}$, these reduced radii $\langle R^2
\rangle^{1/2}/A_C^{1/3}$ show the expected behavior and are larger for \crvi
. So I do not provide a detailed discussion on the intercluster radii $\langle
R^2 \rangle^{1/2}$ because the results are not unique.

\section{Analysis of elastic scattering}
\label{sec:elast}
Finally, it is interesting to test the \al -cluster potentials for \tinull \raa
\tinull\ elastic scattering. Here I first focus on the angular distribution at
$E_\alpha = 25.0$ MeV by Gubler {\it et al.}\ \cite{Gub81} which is available
from the EXFOR database \cite{EXFOR}. This angular distribution covers the
full angular range from about $22^\circ$ to $174^\circ$ in the center-of-mass
system and is thus well-suited for the determination of the optical
potential. In addition, it has been shown for the neighboring \canull \raa
\canull\ scattering that the optical potential at low energies ($E_\alpha = 29$
MeV) is very close to the \al -cluster potential which describes the
\tiiv\ ground state band \cite{Atz96}. For completeness, angular distributions
at 104 MeV \cite{Pesl83} and 140 MeV \cite{Rob78} will also be
analyzed. Unfortunately, elastic \raa\ scattering data are not available for
the radioactive \tiii\ nucleus.

\begin{figure}[htb]
  \includegraphics[width=\columnwidth]{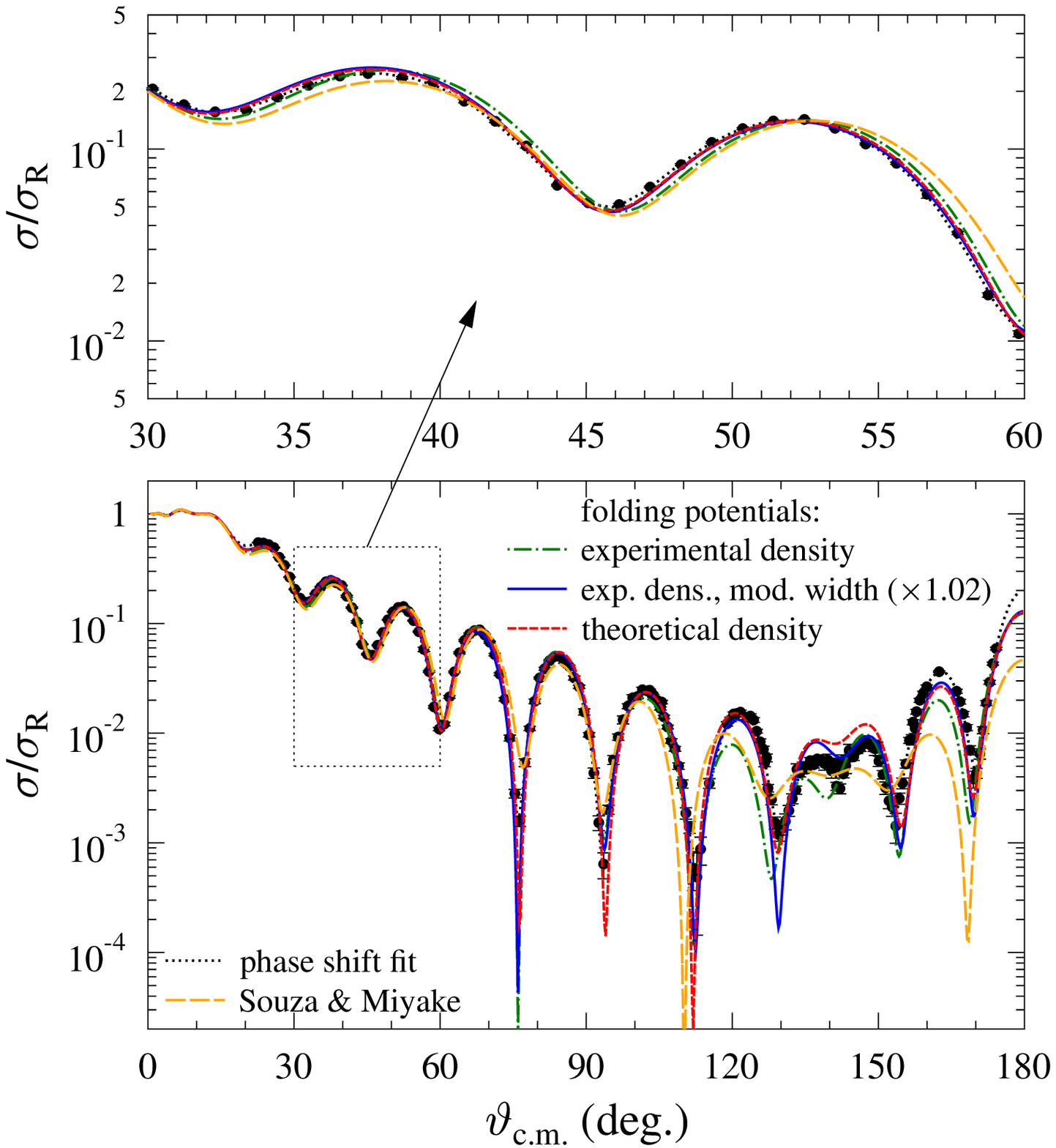}
\caption{
Analysis of the angular distribution of \tinull \raa \tinull\ elastic
scattering at 25 MeV (normalized to Rutherford scattering) \cite{Gub81,EXFOR}:
The phase shift fit (black dotted) reproduces the experimental data almost
perfectly (thus, it is practically invisible behind the experimental data). The
folding potential using the experimental charge density distribution of
\tinull\ (green dash-dotted) is slightly too narrow, resulting in a poor
description of the forward angle diffraction pattern (see enlarged area in the
upper plot); a better description of the experimental data is obtained after
scaling the radial dependence by 2\% (full blue). The folding potential using
the theoretical \tinull\ density (short-dashed red) is able to describe the
experimental data without significant radial modification. The potential of
\SM\ is again too narrow and has a lower $J_R$, leading to deviations in the
forward angle diffraction pattern and at backward angles (long-dashed orange).
}
\label{fig:scat}
\end{figure}

As a first step of the optical model analysis of the \tinull \raa
\tinull\ elastic scattering angular distribution at 25 MeV, a phase shift fit
(according to the technique of \cite{Chi96}) was performed, and it is found
that the experimental angular distribution can be very well reproduced (see
Fig.~\ref{fig:scat}). Next, optical model fits have been performed with the
\al -cluster potential as a starting point for the real part. In extension to
the real potential in Eq.~(\ref{eq:vtot}), a phenomenological potential in the
imaginary part was used which is taken as the sum of volume and surface
Woods-Saxon potentials:
\begin{equation}
W(r) = W_V \times f_{\rm{WS}}(x_V) + W_S \times \frac{df_{\rm{WS}}(x_S)}{dx_S}
\label{eq:imag}
\end{equation}
with
\begin{equation}
f_{WS}(x_{V,S}) = \frac{1}{1+\exp{(x_{V,S})}}
\label{eq:WS}
\end{equation}
and $x_{V,S} = (r-R_{V,S} \times A_T^{1/3})/a_{V,S}$. The depths $W_V$, $W_S$,
radii $R_V$, $R_S$, and diffusenesses $a_V$, $a_S$ of the volume and surface
part have been adjusted to the experimental angular distribution. Note that in
the above definition the maximum depth of the surface part is $-W_S/4$.

A further small modification of the width of the folding potential in the real
part in Eqs.~(\ref{eq:fold}) and (\ref{eq:vtot}) is introduced:
\begin{equation}
V_N(r) = \lambda \, V_F(r/w)
\label{eq:pot_real}
\end{equation}
Here $w$ is a width parameter which should remain very close to unity (see
further discussion below).

Various fits for the angular distribution at 25 MeV \cite{Gub81} are shown in
Fig.~\ref{fig:scat}. The parameters of all fits are listed in Tables
\ref{tab:pot_real} and \ref{tab:pot_imag}.
\begin{table*}
\caption{Parameters of the real part of the optical potential and integral
  values.
}
\label{tab:pot_real}
\begin{center}
\begin{tabular}{rllrrrrrrr}
\hline
\multicolumn{1}{c}{$E_\alpha$} 
& type
& $\rho(^{50}{\rm{Ti}})$
& \multicolumn{1}{c}{$\lambda$}
& \multicolumn{1}{c}{$w$}
& \multicolumn{1}{c}{$J_R$} 
& \multicolumn{1}{c}{$r_{R,rms}$} 
& \multicolumn{1}{c}{$J_I$} 
& \multicolumn{1}{c}{$r_{I,rms}$} 
& \multicolumn{1}{c}{$\sigma_R$}
\\
(MeV)
& & 
& \multicolumn{1}{c}{$-$} 
& \multicolumn{1}{c}{$-$}
& \multicolumn{1}{c}{(MeV\,fm$^3$)} 
& \multicolumn{1}{c}{(fm)}
& \multicolumn{1}{c}{(MeV\,fm$^3$)} 
& \multicolumn{1}{c}{(fm)}
& \multicolumn{1}{c}{(mb)}
\\
%
\hline
%
25.0 
& fold. & exp.~\cite{Vri87}
& 1.246 & $\equiv 1.000$
& 324.9 & 4.305 & 63.5 & 4.580 & 1395
\\
25.0 
& fold. & exp.~\cite{Vri87}
& 1.247 & $1.020$
& 345.1 & 4.391 & 66.8 & 4.409 & 1337
\\
25.0 
& fold. & theo.~\cite{TALYS}
& 1.294 & $0.997$
& 341.8 & 4.388 & 67.4 & 4.448 & 1343
\\
25.0 
& \SM & 
& \multicolumn{1}{r}{$-$} & \multicolumn{1}{r}{$-$}
& 301.1 & 4.323 & 68.5 & 4.646 & 1404
\\
104.0
& fold. & theo.~\cite{TALYS}
& 1.249 & $0.990$
& 289.8 & 4.377 & 88.2 & 5.037 & 1557
\\
104.0
& \SM & 
& \multicolumn{1}{r}{$-$} & \multicolumn{1}{r}{$-$}
& 301.1 & 4.323 & 92.2 & 4.754 & 1413
\\
140.0
& fold. & theo.~\cite{TALYS}
& 1.237 & $1.016$
& 290.3 & 4.505 & 94.9 & 5.275 & 1687
\\
140.0
& \SM\ (V1) & 
& \multicolumn{1}{r}{$-$} & \multicolumn{1}{r}{$-$}
& 301.1 & 4.323 & 239.4 & 4.649 & 1638
\\
140.0
& \SM\ (V2) & 
& \multicolumn{1}{r}{$-$} & \multicolumn{1}{r}{$-$}
& 301.1 & 4.323 & 115.0 & 5.302 & 1805
\\
%
\hline
\end{tabular}
\end{center}
\end{table*}
\begin{table*}
\caption{Parameters of the imaginary part of the optical potential.
}
\label{tab:pot_imag}
\begin{center}
\begin{tabular}{rllrrrrrr}
\hline
\multicolumn{1}{c}{$E_\alpha$} 
& type
& $\rho(^{50}{\rm{Ti}})$
& \multicolumn{1}{c}{$W_V$}
& \multicolumn{1}{c}{$R_V$}
& \multicolumn{1}{c}{$a_V$} 
& \multicolumn{1}{c}{$W_S$} 
& \multicolumn{1}{c}{$R_S$} 
& \multicolumn{1}{c}{$a_S$} \\
(MeV)
& &
& \multicolumn{1}{c}{(MeV)} 
& \multicolumn{1}{c}{(fm)}
& \multicolumn{1}{c}{(fm)}
& \multicolumn{1}{c}{(MeV)} 
& \multicolumn{1}{c}{(fm)}
& \multicolumn{1}{c}{(fm)}
\\
%
\hline
%
25.0 
& fold. & exp.~\cite{Vri87}
& $-19.6$  & 1.385   & 0.263
& $+4.2$   & 1.876   & 0.591
\\
25.0 
& fold. & exp.~\cite{Vri87}
& $-21.7$  & 1.355   & 0.191
& $+9.2$   & 1.741   & 0.391
\\
25.0 
& fold. & theo.~\cite{TALYS}
& $-21.7$  & 1.353   & 0.185
& $+9.7$   & 1.746   & 0.406
\\
25.0 
& \SM & 
& $-20.9$  & 1.386   & 0.220
& $+6.1$   & 1.958   & 0.457
\\
104.0 
& fold. & theo.~\cite{TALYS}
& $-21.0$  & 1.400   & 0.547
& $+14.8$  & 1.631   & 0.608
\\
104.0 
& \SM &
& $-23.1$  & 1.458   & 0.368
& $+12.0$  & 1.691   & 0.459
\\
140.0 
& fold. & theo.~\cite{TALYS}
& $-21.4$  & 1.296   & 0.519
& $+24.5$  & 1.544   & 0.765
\\
140.0 
& \SM\ (V1) &
& $-67.2$  & 1.413   & 0.204
& $+41.0$  & 1.830   & 0.320
\\
140.0 
& \SM\ (V2) &
& $-26.6$  & 2.106   & 0.715
& $-74.6^{\rm{a}}$  & 1.793   & 0.783
\\
%
\hline
\multicolumn{9}{l}{\footnotesize{$^{\rm{a}}$Note that the sum of volume and
    surface imaginary potential does not change its sign.}} \\
\end{tabular}
\end{center}
\end{table*}

The folding potential which is based on the experimental charge density
distribution of \tinull\ does not exactly match the diffraction pattern in the
forward region. The maxima and minima of the theoretical angular distribution
are shifted towards larger angles; this corresponds to a potential which is
slightly too narrow. After scaling the radial dependence of the real potential
by a factor of $ w = 1.02$ a much better fit is obtained (for technical
details, see \cite{Mohr13}). Such a behavior is not surprising for
\tinull\ with $N/Z \approx 1.27$ because the neutron distribution should have
a larger radial extent than the proton (charge) distribution. Evidence for a
larger radial extent of the neutron distribution of \tinull\ was also derived
from \al\ scattering at higher energies \cite{Gils84}.

Fortunately it turns out that the folding potential which is calculated from
the theoretical density distribution of \tinull\ (as provided in TALYS) is
able to reproduce the experimental angular distribution without a significant
radial scaling (the best fit is obtained using a scaling parameter of $w =
0.997$, i.e., very close to unity). This confirms the reliability of the
theoretical density distribution of \tinull\ which is consequently used
throughout this paper.

The best-fit potential shows a real volume integral of $J_R \approx 342$
MeV\,fm$^3$ and $r_{R,rms} \approx 4.39$ fm; the imaginary part has $J_I
\approx 67$ MeV\,fm$^3$ and $r_{I,rms} \approx 4.45$ fm. (As usual, the
negative sign of the volume integrals is negelected in the discussion of $J_R$
and $J_I$.) The real volume integral $J_R$ from elastic scattering is close to
the results from scattering on neighboring nuclei \cite{Atz96} but
significantly larger than the values for the \al -cluster states which are
around $310 - 315$ MeV\,fm$^3$ (see Table \ref{tab:res_iv}). For very
pronounced \al -cluster nuclei like \tiiv\ or $^{20}$Ne such deviations are
neither expected nor found \cite{Atz96,Abe93}, and thus this deviation for
\criv\ may be interpreted that \criv\ is not a very pure \tinull\ $\otimes$
\al\ cluster. This interpretation is in agreement with the smaller reduced
widths which are found for \criv\ (in comparison to \crvi\ and \tiiv ).

As the potential by \SM\ has been optimized for the description of \al
-cluster states with low $J_R \approx 300$ MeV\,fm$^3$, it is not surprising
that this potential cannot reproduce the \tinull \raa \tinull\ angular
distribution with similar quality as the double-folding
potentials. Furthermore, $r_{R,rms} = 4.32$ fm of the \SM\ potential is also
lower than the best-fit values around 4.39 fm by about 2\%. 

At higher energies the differences between the folding potential and the
potential by \SM\ become even more obvious. Indeed, the volume integral $J_R$
and the $rms$ radius of the \SM\ potential are close to the results of
Roberson {\it et al.}\ at $E_\alpha = 140$ MeV \cite{Rob78}. However, angular
distributions at higher energies are not only sensitive to the potential at
the nuclear surface but depend on the shape of the potential in a wider radial
range. In the following I analyze the angular distributions at 140 MeV by
Roberson {\it et al.}\ \cite{Rob78} and at 104 MeV by Pesl {\it et
  al.}\ \cite{Pesl83}. Unfortunately, the 140 MeV data (as provided by EXFOR)
had to be re-read from Fig.~1 of \cite{Rob78}; this leads to increased
uncertainties and lower reliability of the extracted parameters. As no
uncertainties are provided in EXFOR, a fixed uncertainty of 5\% was assumed
for the fitting procedure. Furthermore, the angular range of the 140 MeV data
is limited to about $75^\circ$, and the number of data points (51) is not very
high. The angular distribution at 104 MeV \cite{Pesl83} covers a larger
angular range up to $90^\circ$, consists of a factor of three more data
points, and is available numerically (including uncertainties) from the
internal report KFK-3242 \cite{Pesl82}.

\begin{figure}[htb]
  \includegraphics[width=\columnwidth]{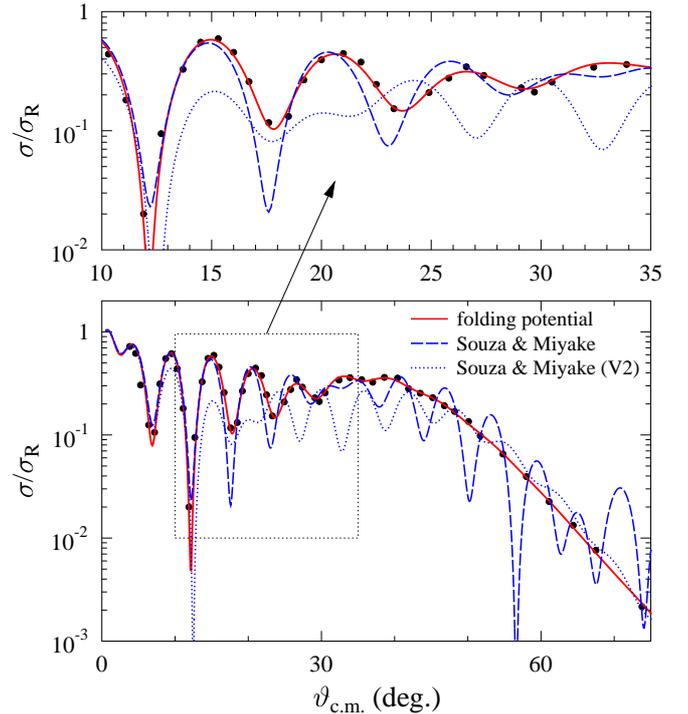}
\caption{ 
Analysis of the angular distribution of \tinull \raa \tinull\ elastic
scattering at 140 MeV (normalized to Rutherford scattering)
\cite{Rob78,EXFOR}. The experimental data are nicely reproduced using a
folding potential (full red line) whereas two different fits (dashed and
dotted blue lines) with the fixed \SM\ potential in the real part show larger
deviations. Further discussion see text.
}
\label{fig:scat140}
\end{figure}

The 140 MeV data can be nicely reproduced by the folding potential (see
Fig.~\ref{fig:scat140}). The resulting volume integrals $J_R$ and $J_I$ are
close to the results obtained for neighboring nuclei \cite{Atz96}: $J_R$
decreases with increasing energy, and $J_I$ approaches a saturation value of
about 100 MeV\,fm$^3$.  An early study with folding potentials has also
determined similar volume integrals \cite{Kob84b}. Thus, the present analysis
mainly confirms that the chosen theoretical density of \tinull\ in combination
with the DDM3Y interaction is appropriate.

Keeping a fixed real part from \SM\ and adjusting only the imaginary potential
leads to a significantly worse reproduction of the experimental angular
distribution. The best fit shows a strongly oscillating angular distribution
at backward angles, correlated with very different imaginary volume integral
$J_I$. It is obvious from Fig.~\ref{fig:scat140} that the strongly oscillating
backward cross sections are not sufficiently constrained by the few
experimental data points above about $50^\circ$. A further local minimum in
$\chi^2$ can be found with reasonable volume integrals (V2 in
Fig.~\ref{fig:scat140} and Tables \ref{tab:pot_real} and \ref{tab:pot_imag});
however, this results in a relatively poor reproduction of the experimental
data at forward angles.

\begin{figure}[htb]
  \includegraphics[width=\columnwidth]{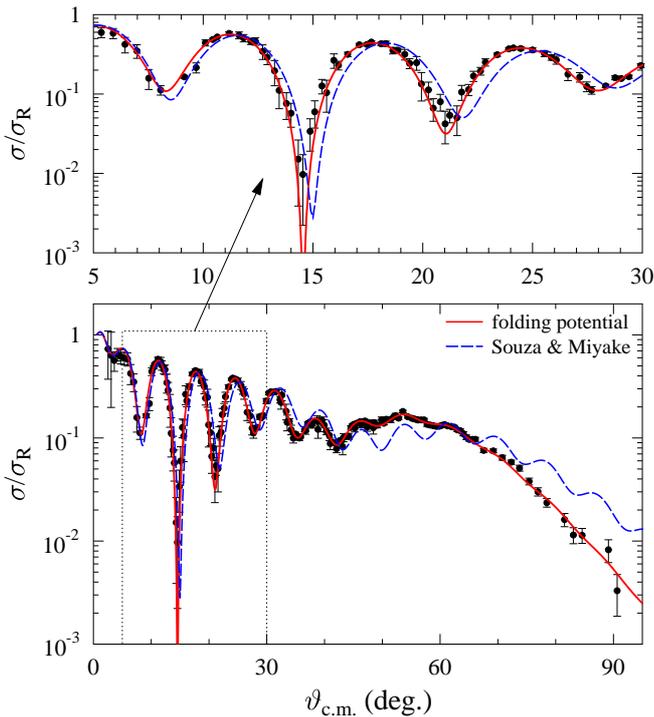}
\caption{
Analysis of the angular distribution of \tinull \raa \tinull\ elastic
scattering at 104 MeV (normalized to Rutherford scattering)
\cite{Pesl83,EXFOR}. Similar to the 140 MeV data, the folding potential (full
red line) reproduces the experimental angular distribution, but the
\SM\ potential (dashed blue line) shows larger deviations. Further discussion
see text.
}
\label{fig:scat104}
\end{figure}

Also the angular distribution at 104 MeV \cite{Pesl83} is very well reproduced
by the folding potential (see Fig.~\ref{fig:scat104}). Interestingly, the
width parameter for the 104 MeV data is very close (within 0.7\%) to the 25
MeV fit. For the 140 MeV data the resulting width parameter was about 2\%
higher. This finding may have been affected from the uncertainties of the
re-digitization of the 140 MeV data and the missing experimental
uncertainties.

The fit with the fixed \SM\ real part results in volume integrals which are
very close to the folding potential results. It is not surprising that the
fixed \SM\ real part cannot reproduce the experimental angular distribution
with the accuracy as the folding potential with two adjustbale parameters
(strength $\lambda$ and width $w$). But even if the same scaling parameters
($\lambda$ and $w$) are applied to the \SM\ potential, it is not possible to
reproduce the angular distribution with a similar quality as the folding
potential.
 
Summarizing the above analysis of elastic scattering angular distributions, it
can be stated that the folding potential is able to describe the angular
distributions over a wide energy range with smoothly varying parameters. The
width parameter $w$ remains very close to unity using a theoretical density of
\tinull\ whereas a folding potential derived from the experimental charge
distribution requires a radial scaling of about 2\%. 

The potential of \SM\ is close to the folding potential whose parameters have
been adjusted to elastic scattering angular distributions. This is a quite
noticeable finding because the \SM\ potential was optimized only for the
reproduction of excitation energies of \criv . Without any further adjustment
of the real part, the \SM\ potential cannot reach the same accuracy in the
reproduction of the angular distributions, but it still provides a reasonable
description of the experimental data at 25 MeV and 104 MeV.

\section{Summary and conclusions}
\label{sec:conc}
\al -cluster states in \crvi\ and \criv\ are studied on the basis of a
double-folding potential. A minor and smooth variation of the potential
strength is found for the states under study in \crvi\ and \criv . Reduced
widths, intercluster separations, and transition strengths are close to the
results of a recent study by Souza and Miyake \cite{Sou17} using a specially
shaped potential, and the calculated transition strengths show reasonable
agreement with the experimental values without effective charges. The present
study confirms the significant \al -cluster properties of \crvi\ and smaller
reduced widths for \criv . Furthermore, it is found that the specially shaped
potential by \SM\ is close to the double-folding potential of the present
study (except at very small radii).

The application of the double-folding potential to elastic \tinull \raa
\tinull\ scattering provides excellent fits for the angular distributions over
a wide energy range from 25 MeV to 140 MeV. The derived parameters show that
the underlying \tinull\ density should be preferentially taken from
theory. The potential which is calculated from the experimental charge density
distribution, turns out to be too narrow by about 2\%, thus indicating a
larger radial extent of the neutron distribution in \tinull .

\section*{Acknowledgments}
I thank M.\ A.\ Souza and H.\ Miyake for providing their results amd for
encouraging discussions. This work was supported by OTKA (K108459 and K120666).



\begin{thebibliography}{}
%
\bibitem{Sou17}
M.\ A.\ Souza and H.\ Miyake,
Europ.\ Phys.\ J.\ A {\bf 53}:146 (2017)
%
\bibitem{Hor12}
H.\ Horiuchi, K. Ikeda, K.\ Kat{\={o}},
Prog.\ Theor.\ Phys.\ Suppl.\ {\bf 192}, 1 (2012).
%
\bibitem{Ohk98}
S.\ Ohkubo, M.\ Fujiwara, P.\  E.\ Hodgson,
Prog.\ Theor.\ Phys.\ Suppl.\ {\bf 132}, 1 (1998).
%
\bibitem{Mic98}
F.\ Michel, S.\ Ohkubo, G.\ Reidemeister,
Prog.\ Theor.\ Phys.\ Suppl.\ {\bf 132}, 7 (1998).
%
\bibitem{Yam98}
T.\ Yamaya, K.\ Katori, M.\ Fujiwara, S.\ Kato, S.\ Ohkubo,
Prog.\ Theor.\ Phys.\ Suppl.\ {\bf 132}, 73 (1998).
%
\bibitem{Sak98}
T.\ Sakuda and S.\ Ohkubo,
Prog.\ Theor.\ Phys.\ Suppl.\ {\bf 132}, 103 (1998).
%
\bibitem{Ueg98}
E.\ Uegaki,
Prog.\ Theor.\ Phys.\ Suppl.\ {\bf 132}, 135 (1998).
%
\bibitem{Has98}
M.\ Hasegawa,
Prog.\ Theor.\ Phys.\ Suppl.\ {\bf 132}, 177 (1998).
%
\bibitem{Koh98}
S.\ Koh,
Prog.\ Theor.\ Phys.\ Suppl.\ {\bf 132}, 197 (1998).
%
\bibitem{Toh98}
A.\ Tohsaki,
Prog.\ Theor.\ Phys.\ Suppl.\ {\bf 132}, 213 (1998).
%
\bibitem{Des02}
P.\ Descouvemont,
Nucl.\ Phys.\ {\bf A709}, 275 (2002).
%
\bibitem{Sak02}
T.\ Sakuda and S.\ Ohkubo,
Nucl.\ Phys.\ {\bf A712}, 59 (2002).
%
\bibitem{Mohr93}
P.\ Mohr, H.\ Abele, V.\ K\"olle, G.\ Staudt, H.\ Obrhummer, H.\ Krauss,
Z.\ Phys.\ A {\bf 349}, 339 (1993).
%
\bibitem{Abe93}
H.\ Abele and G.\ Staudt,
Phys.\ Rev.\ C {\bf 47}, 742 (1993).
%
\bibitem{Wil02}
S.~Wilmes, V.~Wilmes, G.~Staudt, P.~Mohr, J.~W.~Hammer,
Phys.\ Rev.\ C {\bf {66}}, 065802 (2002).
%
\bibitem{Mohr05}
P.\ Mohr,
Phys.\ Rev.\ C {\bf 72}, 035803 (2005).
%
\bibitem{Mohr06}
P.\ Mohr, C.\ Angulo, P.\ Descouvemont, H.\ Utsunomiya,
Europ.\ Phys.\ J.\ A 27, s01, 75 (2006).
%
\bibitem{Atz96} 
  U.\ Atzrott, P.\ Mohr, H.\ Abele, C.\ Hillenmayer, and
  G.\ Staudt,
  Phys.\ Rev.\ C {\bf 53}, 1336 (1996).
%
\bibitem{Hoy94}
F.\ Hoyler, P.\ Mohr, G.\ Staudt,
Phys.\ Rev.\ C {\bf 50}, 2631 (1994).
%
\bibitem{Ohk95}
S.\ Ohkubo,
Phys.\ Rev.\ Lett.\ {\bf 74}, 2176 (1995).
%
\bibitem{Mohr07}
P.\ Mohr,
Europ.\ Phys.\ J.\ A {\bf 31}, 23 (2007).
%
\bibitem{Ohk09}
S.\ Ohkubo,
Int.\ J.\ Mod.\ Phys.\ A {\bf 14}, 2035 (2009).
%
\bibitem{Mohr08}
P.\ Mohr,
The Open Nuclear and Particle Physics Journal {\bf 1}, 1 (2008).
%
\bibitem{Mohr00}
P.\ Mohr,\ Phys.\ Rev.\ C {\bf 61}, 045802 (2000).
%
\bibitem{Xu06}
Chang Xu and Zhongzhou Ren,
Phys.\ Rev.\ C {\bf 73}, 041301(R) (2006).
%
\bibitem{Mohr17}
P.\ Mohr,
Phys.\ Rev.\ C {\bf 95}, 011302(R) (2017).
%
\bibitem{Sat79}
G.\ R.\ Satchler and W.\ G.\ Love,
Phys.\ Rep.\ {\bf 55}, 183 (1979).
%
\bibitem{Kob84}
A.\ M.\ Kobos, B.\ A.\ Brown, R.\ Lindsay, and G.\ R.\ Satchler,
Nucl.\ Phys.\ {\bf A425}, 205 (1984).
%
\bibitem{Mohr13}
P.\ Mohr, G.\ G.\ Kiss, Zs.\ F\"ul\"op, D.\ Galaviz, Gy.\ Gy\"urky, 
E.\ Somorjai,
At.\ Data Nucl.\ Data Tables {\bf 99}, 651 (2013).
%
\bibitem{Vri87}
H.\ de Vries, C.\ W.\ de Jager, and C.\ de Vries,
        Atomic Data and Nuclear Data Tables {\bf 36}, 495 (1987).
%
\bibitem{Tsu17}
K.\ Tsukuda {\it et al.},
Phys.\ Rev.\ Lett.\ {\bf 118}, 262501 (2017).
%
\bibitem{TALYS}
A.\ J.\ Koning, S.\ Hilaire, S.\ Goriely,
computer code TALYS, version 1.8,
{\it{http://www.talys.eu}};
A.\ J.\ Koning, S.\ Hilaire, and M.\ C.\ Duijvestijn,
AIP Conf.\ Proc.\ \textbf{769}, 1154 (2005).
%
\bibitem{Gub81}
H.\ P.\ Gubler, U.\ Kiebele, H.\ O.\ Meyer, G.\ R.\ Plattner, I.\ Sick,
Nucl.\ Phys.\ {\bf A351}, 29 (1981).
%
\bibitem{Rob78}
P.\ L.\ Roberson, D.\ A.\ Goldberg, N.\ S.\ Wall, L.\ W.\ Woo, H.\ L.\ Chen,
Phys.\ Rev.\ Lett.\ {\bf 42}, 54 (1978).
%
\bibitem{Pesl83}
R.\ Pesl, H.\ J.\ Gils, H.\ Rebel, E.\ Friedman, J.\ Buschmann,
H.\ Klewe-Nebenius, S.\ Zagromski,
Z.\ Phys.\ A {\bf 313}, 111 (1983).
%
\bibitem{ENSDF}
Online database ENSDF,
available at {\it{www.nndc.bnl.gov/ensdf}};
based on \cite{NDS46,NDS54}.
%
\bibitem{NDS46}
S.-C.\ Wu,
Nucl.\ Data Sheets {\bf 91}, 1 (2000).
%
\bibitem{NDS54}
Yang Dong and Huo Junde,
Nucl.\ Data Sheets {\bf 121}, 1 (2014).
%
\bibitem{Buck95}
B.\ Buck, A.\ C.\ Merchant, S.\ M.\ Perez,
Phys.\ Rev.\ C {\bf 51}, 559 (1995).
%
\bibitem{Mic83}
F.\ Michel, J\ Albinski, P.\ Belery, Th.\ Delbar, Gh.\ Gr{\'e}goire,
B.\ Tasiaux, G.\ Reidemeister,
Phys.\ Rev.\ C {\bf 28}, 1904 (1983).
%
\bibitem{Mic00}
F.\ Michel, G.\ Reidemeister, S.\ Ohkubo,
Phys.\ Rev.\ C 61, 041601(R) (2000).
%
\bibitem{McF66}
L.\ McFadden and G.\ R.\ Satchler, 
Nucl.\ Phys.\ {\bf 84}, 177 (1966).
%
\bibitem{Sou15}
M.\ A.\ Souza and H.\ Miyake,
Phys.\ Rev.\ C {\bf 91}, 034320 (2015).
%
\bibitem{Mic88}
F.\ Michel, G.\ Reidemeister, S.\ Ohkubo,
Phys.\ Rev.\ C {\bf 37}, 292 (1988).
%
\bibitem{Buck75}
B.\ Buck, C.\ B.\ Dover, J.\ P.\ Vary,
Phys.\ Rev.\ C {\bf 11}, 1803 (1975).
%
\bibitem{EXFOR}
N.\ Otuka {\it et al.},
Nucl.\ Data Sheets {\bf 120}, 272 (2014);
EXFOR database available online at
{\it{http://www-nds.iaea.org/exfor/exfor.htm}}.
%
\bibitem{Chi96}
V.\ Chist{\'e}, R.\ Lichtenth{\"a}ler, A.\ C.\ C.\ Villari, L.\ C.\ Gomes,
Phys.\ Rev.\ C {\bf 54}, 784 (1996).
%
\bibitem{Gils84}
H.\ J.\ Gils, H.\ Rebel, E.\ Friedman,
Phys.\ Rev.\ C {\bf 29}, 1295 (1984).
%
\bibitem{Pesl82}
R.\ Pesl,
internal report KFK-3242, Kernforschungszentrum Karlsruhe (1982); available at
{\it{https://publikationen.bibliothek.kit.edu/270016791}}.
%
\bibitem{Kob84b}
A.\ M.\ Kobos, B.\ A.\ Brown, P.\ E.\ Hodgson, G.\ R.\ Satchler,
A. Budzanowski,
Nucl.\ Phys.\ {\bf A384}, 65 (1982).
%
\end{thebibliography}
\end{document}